\documentclass[reprint,aps,final,longbibliography,superscriptaddress,nobibnotes,footinbib,amsfonts,showpacs,letterpaper,jmp,amssymb,pdftex]{revtex4-1}
\usepackage{graphicx,multirow,bm,amsmath}
\usepackage{booktabs}
\usepackage{hyperref}
\usepackage{bbm}
\usepackage[usenames,dvipsnames]{color}
\usepackage[table,xcdraw]{xcolor}

\newcommand{\kbrnasaames}{KBR, Inc., Intelligent Systems Division, NASA Ames Research Center, Moffett Field, CA 94035, USA}
\newcommand{\nasaames}{Intelligent Systems Division, NASA Ames Research Center, Moffett Field, CA 94035, USA}

\begin{document}

	\title{Thermodynamic Integration for Dynamically Unstable Systems Using Interatomic Force Constants without Molecular Dynamics}
	\author{Junsoo Park}
	\email{junsoo.park@nasa.gov}
	\affiliation{\kbrnasaames}
    \author{Zhigang Wu}
    \affiliation{\nasaames}
    \author{John W. Lawson}
	\affiliation{\nasaames}
	\date{\today} 
	\begin{abstract}
We demonstrate an efficient and accurate, general-purpose first-principles blueprint for calculating anharmonic vibrational free energy and predicting structural phase transition temperatures of solids. Thermodynamic integration is performed without molecular dynamics using only interatomic force constants to model analogues of the true potential and generate their thermal ensembles. By replacing \textit{ab initio} molecular dynamics (AIMD) with statistical sampling of ensemble configurations and trading density-functional theory (DFT) energy calculations on each configuration for a set of matrix operations, our approach enables a faster thermodynamic integration by 4 orders of magnitude over the traditional route via AIMD. Experimental phase transition temperatures of a variety of strongly anharmonic materials with dynamical instabilities including shape-memory alloys are recovered to largely within $25$\% error. Such a combination of speed and accuracy enables the method to be deployed at a large-scale for predictive mapping of phase transition temperatures.
	\end{abstract}
	\maketitle

Structural phase transition is a fundamental behavior of materials and directly responsible for important functionalities such as ferroelectricity and shape-memory. Because different phases of the same chemical composition often deliver notably different sets of properties, structural phase transition also serves as an important engineering design consideration for an even wider gamut of functional applications. Phase transition is driven by relative free energies of phases in contention at given external conditions such as temperature and pressure. At any given condition, the phase with the lowest free energy is the most thermodynamically stable and thus, barring kinetic effects, manifest in real life. While there are many potential contributors of free energy in solids (e.g. vibrational, configurational, electronic), the vibrational part dominates over others in most cases. The harmonic part of vibrational free energy derives straightforwardly from the theory of phonons, but the anharmonic part is much more difficult to define or compute accurately.

Thermodynamic integration (TI), also known as $\lambda$-integration, is an exact and general theoretical definition for free energy difference (or correction) between any two potentials. It states that
\begin{equation}\label{eq:ti}
\Delta F = \int_{0}^{1} \left\langle \frac{\partial U_{\lambda}}{\partial \lambda} \right\rangle d\lambda
\end{equation}
where the potential energy $U_{\lambda}$ is a $\lambda$-modulated mixture of, as usually implemented, true and some approximate reference potential energies as $U_{\lambda}=\lambda U_{\text{true}}+(1-\lambda)U_{\text{approx}}$. It follows then $\langle \frac{\partial U_{\lambda}}{\partial \lambda} \rangle=\langle U_{\text{true}}-U_{\text{approx}} \rangle$, where the bracket denotes an ensemble average over configurations generated from the potential. In the context of first-principles calculation for anharmonic free energy, $U_{\text{true}}$ is the Born-Oppenheimer (BO) potential given by density-functional theory (DFT) \cite{hohenbergkohn,kohnsham} while $U_{\text{approx}}$ is, usually, some effective temperature-dependent harmonic phonon representation of the true potential. When put to use in conjunction with \textit{ab initio} molecular dynamics (AIMD) powered by DFT, as traditionally done, TI-AIMD represents the most accurate method for computing $\Delta F$. It has accurateluy described phase transitions in strongly anharmonic materials including some of the most pathological systems, shape-memory alloys (SMAs)\cite{tiaimdnitiHaskins,tiaimdnitiHaskins2,tiaimdbinarysmaHaskins,tiaimdnitiZW,tiaimdnitioffZW,tiaimdnitihfZW}. In spite of its rigor and generality, this approach is limited by severe computational demand owing to 1) AIMD over several picoseconds for each $\lambda$ to sample sufficient number of ensemble configurations (routinely on the order of $10^{3}$) for converged ensemble average; 2) accurate DFT energy calculation for each configuration sampled. When performed over a grid of $\lambda$ and temperature, the computation easily amounts to millions of CPU hours to become quickly prohibitive. An alternative that has gained traction is MD with machine-learned force field (MD-MLFF) \cite{mdmlffzr,mdmlffzro21,mdmlffzro22,mdmlffgete}, but this also requires some AIMD run for on-the-fly force-field training and has demanded CPU hours on the order of $10^5$.

A more expedient route to anharmonic free energy is offered by interatomic force constants (IFC) and corrections based on perturbation theory. It starts from the Taylor expansion of the BO lattice potential energy with respect to atomic displacements $u$:
\begin{widetext}
\begin{equation}\label{eq:latticeenergy}
U=U_{0}+\frac{1}{2!}\sum_{ab}\Phi_{2}^{ab}u^{a}u^{b}+\frac{1}{3!}\sum_{abc}\Phi_{3}^{abc}u^{a}u^{b}u^{c}+\frac{1}{4!}\sum_{abcd}\Phi_{4}^{abcd}u^{a}u^{b}u^{c}u^{d}+\cdots.
\end{equation}
\end{widetext}
Here, $\Phi_{n}$ denotes $n$th-order IFC quantifying the interaction strength of an $n$-body cluster of atoms in their respective Cartesian displacement directions, (collectively denoted by $a$, $b$, etc.). The leading term is of second order, describing pair-wise harmonic interaction, i.e. phonons, whose free energy is straightforward to compute. Owing to the developments of the past decade \cite{csld1,csld2,csld3,alamode,alamodereview,hiphive,hiphivetest}, accurate and very efficient calculation of IFC up to high orders by fitting them to a set of DFT training forces has now become routine. If a material exhibits imaginary phonon modes, then dynamically stable phonons at finite temperatures can be computed using one of the many methods developed for such a purpose \cite{tdep1,tdep2,tdep3,sscha1,sscha2,scaild1,scaild2,qscaild1,qscaild2,aprnpbte,aprngete,specialdisplacement}. Anharmonic free energies due to 3rd and 4th order IFCs have been derived from many-body perturbation theory and thermodynamic Green's functions \cite{cowleylatticedynamics,improvedsc1968,allenlatticedynamics,sscha3,sscha4,alamodesrtio3,alamodecspbbr3} and pose substantially lighter computational demand than TI-AIMD. These corrections have been successfully applied to many systems \cite{alamodesrtio3,alamodecspbbr3,alamodescf3,alamodebatio3,alamodeyco5,sschasnse}, but have been unsuccessful for other more strongly anharmonic systems, notably ZrO$_{2}$ \cite{alamodezro2}. By design, the validity of perturbation theory is confined to the regime where the perturbation (anharmonicity) is relatively weak. 

\begin{figure}
\includegraphics[width=1\linewidth]{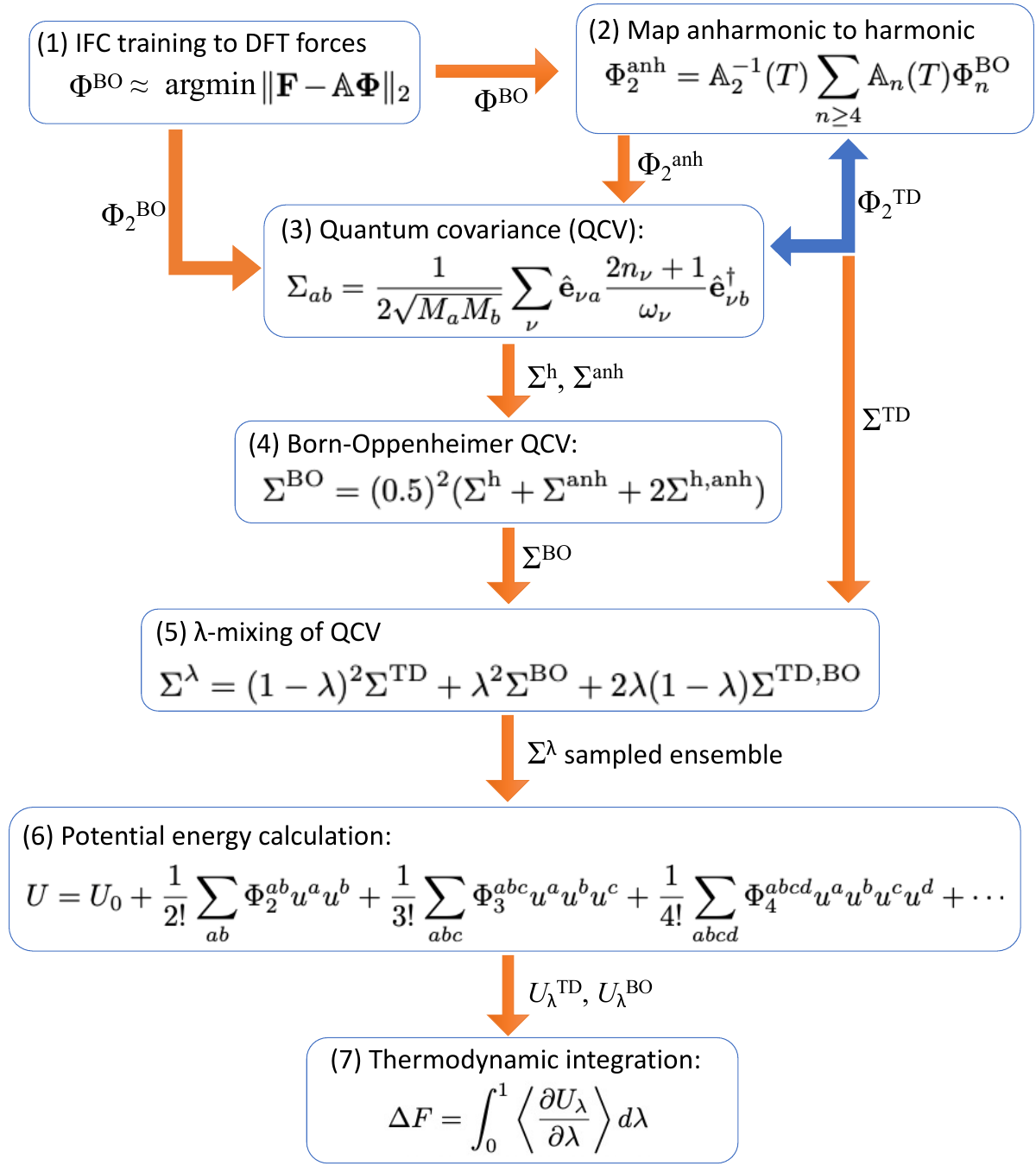}
\caption{\textbf{The flowchart of TI-IFC.} Key steps, equations and transfer of variables in between are listed.}
\label{fig:flowchart}
\end{figure}

There is a clear need for a computationally efficient route to anharmonic free energy that pays minimal penalty in the theoretical rigor and general validity of TI-AIMD \cite{freeenergymethodsassessment,freeenergypredictionsforcrystalstability}. We hereby present one such method that performs TI using only IFC (TI-IFC). In short, we combine anharmonic IFC and temperature-dependent harmonic IFC to statistically sample $\lambda$-dependent thermal ensembles and calculate their potential energies directly with the IFCs. The flowchart for TI-IFC is shown in Fig. \ref{fig:flowchart}. This approach requires no AIMD simulation at all and no further DFT calculation beyond what is required of the original IFC-fitting, enabling 4 orders of magnitude speed-up in computation over TI-AIMD. We validate its performance across various materials exhibiting dynamical instabilities, notably binary SMAs, with experimentally known phase transitions as well as previous computational studies using AIMD or MD-MLFF.


\begin{figure*}
\includegraphics[width=1\linewidth]{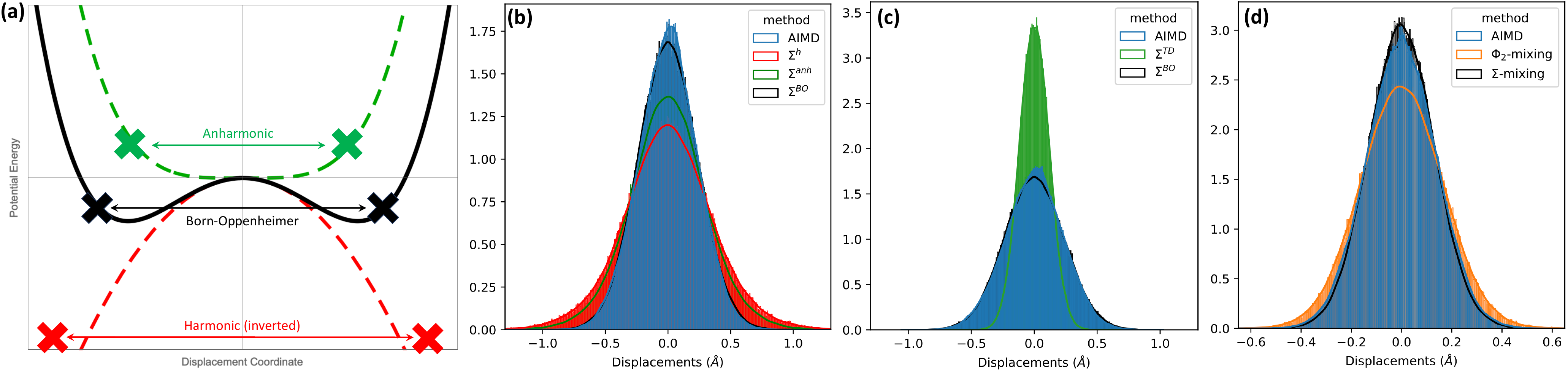}
\caption{\textbf{Statistical sampling of thermal ensemble.} \textbf{a)} Schematic of a ``double-well" potential as a whole (solid black) and decomposed into its inverted-harmonic (dashed red) and 4th-order anharmonic (dashed green) parts. The horizontal arrows and the bolded ``X" marks indicate the hypothetical spread of the atomic displacements under each potential. It depicts that the true ensemble could be statistically sampled by incorporating the distributions from the inverted harmonic and anharmonic potentials whereby the latter is approximated as an upright harmonic by $\Phi_{2}^{\rm anh}$ in Eq. 3. \textbf{b-d)} Comparison of displacement profiles of B2-PdTi ensembles at 800 K using $4\times4\times4$ supercell and 6th-order expansion. \textbf{b)} Ensemble sampled from $\Sigma^{\text{BO}}$ (black), constructed by mixing harmonic (\color{red}red\color{black}) and anharmonic (\color{green}green\color{black}) potentials, approximates the AIMD profile (\color{blue}blue\color{black}) accurately. \textbf{c)} $\Sigma^{\text{BO}}$-sampling approximates AIMD displacement profile much better than $\Sigma^{\text{TD}}$-sampling, which corresponds to $\lambda=0$ in the context of TI. \textbf{d)} $\Sigma^{\lambda}$-sampling better approximates AIMD profile (\color{blue}blue\color{black}) than straight-up $\Phi_{2}$-mixing (\color{orange}orange\color{black}).}
\label{fig:displacements}
\end{figure*}


The first prerequisite is IFCs up to at least 4th order. 4th-order anharmonicity is the first order capable of righting the inverted harmonic potential of dynamically unstable systems by rendering a so-called ``double-well" or ``sombrero" potential. We obtain IFCs by fitting them to DFT forces on a handful of randomly displaced training configurations. Let us denote these IFCs $\Phi^{\text{BO}}$ as they are meant to desecribe the BO potential. We use the HiPhive package \cite{hiphive} and recursive feature elimination for training IFC. All DFT calculations for structural relaxation and IFC-fitting are performed with Vienna \textit{Ab initio} Simulations Package (VASP) \cite{vasp1,vasp2,vasp3,vasp4} with the Perdew-Burke-Ernzerhof exchange-correlation functional \cite{pbe} modified for solids (PBEsol) \cite{pbesol} and projector-augmented wave (PAW) pseudopotentials \cite{paw,ultrasofttopaw}. All lattice dynamic calculations are performed using PBEsol lattice and IFC without exception.

The second prerequisite is temperature-dependent effective harmonic IFC ($\Phi_{2}^{\text{TD}}$), which is obtained via a routine that closely follows that of Refs. \onlinecite{aprnpbte,aprngete}. It starts by redistributing the anharmonic forces to pair interactions resulting in artificially harmonic (2nd-order) IFCs that approximate the anharmonic profile, 
\begin{equation}\label{eq:phianharmonic}
\Phi_{2}^{\text{anh}} = \mathbb{A}_{2}^{-1}(T) \sum_{n\ge4}\mathbb{A}_{n}(T)\Phi_{n}^{\text{BO}},
\end{equation}
which is then added to and renormalizes $\Phi_{2}^{\text{BO}}$ so that 
\begin{equation}\label{eq:renormalization}
\Phi_{2}^{\text{TD}} = \Phi_{2}^{\text{BO}} + \Phi_{2}^{\text{anh}}.
\end{equation}
Temperature-dependence enters in the form of displacement matrices $\mathbb{A}$ constructed from the thermal ensemble assembled by Gaussian random sampling ($\mathbf{u}\sim N(0,\Sigma)$). The key quantity is the quantum covariance (QCV) matrix \cite{qscaild1,qscaild2} of displacements, which is calculable for any harmonic IFC between all pairs of atoms in their respective Cartesian directions, including quantum effect, as
\begin{equation}\label{eq:qcv}
\Sigma_{ab} = \frac{1}{2\sqrt{M_{a}M_{b}}} \sum_{\nu}  \hat{\mathbf{e}}_{\nu a} \frac{2n_{\nu}+1}{\omega_{\nu}}  \hat{\mathbf{e}}_{\nu b}^{\dagger},
\end{equation}
where $\hat{\mathbf{e}}$ are the eigenvectors, $\omega$ is the frequency, and $n$ is the temperature-dependent Bose-Einstein population for phonon mode $\nu$, all obtained from phonon diagonalization. Eqs. \ref{eq:phianharmonic}$\sim$\ref{eq:qcv} are iterated until convergence.


To perform TI, one needs physically reasonable thermal ensembles at each $\lambda$, which has been traditionally sampled with MD at large computational costs. We instead venture to statistically sample thermal ensembles in order to substantially reduce compute time while guaranteeing practically zero correlation between configurations. The all-critical problem is to implement it in a way that ensures the physicality of ensembles that adhere to the true BO potential at $\lambda\rightarrow1$ without deferring to MD.

A tempting way forward may be to directly mix the harmonic IFCs as $\Phi_{2}^{\lambda} =(1-\lambda)\Phi_{2}^{\text{TD}} +\lambda\Phi_{2}^{\text{BO}}$, with which a QCV could be straightforwardly constructed by Eq. \ref{eq:qcv}. Viable as this may be if no imaginary modes are involved, for dynamically unstable phases, $\Phi^{\lambda}_{2}$ would exhibit imaginary phonons toward $\lambda\rightarrow 1$. $\Phi^{\lambda}_{2}$ does not account for any anharmonic interactions to right an inverted harmonic well, so sampling with the QCV from $\Phi^{\lambda}_{2}$ would yield unphysically large atomic displacements far past the local minima of the true BO potential that AIMD would correctly sample. Refer to Fig. \ref{fig:displacements}a for a graphical explanation. In the presence of an anharmonic (e.g. fourth-order) interaction, true atomic displacements would be distributed intermediate the inverted harmonic well and the anharmonic well near the local minima. 

To resolve this problem, we devise a statistically inspired method that performs $\lambda$-mixing at the level of not $\Phi_{2}$ but $\Sigma$. The core idea is to treat displacements owing to the harmonic and anharmonic terms as two independent sets of random variables and mix their underlying distributions. The starting point is none other than $\Phi_{2}^{\text{anh}}$ from Eq. \ref{eq:phianharmonic}, the anharmonic interactions reduced as harmonics. Its QCV, $\Sigma^{\text{anh}}$, approximates the distribution of anharmonic displacements. The QCV of the original harmonic term is $\Sigma^{\text{h}}$. We then apply the statistical theory of covariance of a weighted sum of two independent variables to theorize the QCV abiding by the full BO potential as
\begin{equation}\label{eq:qcvdft}
\Sigma^{\text{BO}} = (0.5)^{2}(\Sigma^{\text{h}}+\Sigma^{\text{anh}}+2\Sigma^{\text{h,anh}})
\end{equation} 
where $\Sigma^{\text{h,anh}}=\mathbb{E}\left[\left(\mathbf{u}^{\text{h}}-\mathbb{E}\left[\mathbf{u}^{\text{h}}\right]\right)\left(\mathbf{u}^{\text{anh}}-\mathbb{E}\left[\mathbf{u}^{\text{anh}}\right]\right)\right]$ is the cross-covariance, and $\mathbb{E}$ represents the expected value. The weights are 0.5 for both $\Sigma^{\text{h}}$ and $\Sigma^{\text{anh}}$ since the two terms must enter in equal weights. $\Sigma^{\text{BO}}$ constructed as such yields an ensemble that approximates the AIMD ensemble very well, as shown in Figs. \ref{fig:displacements}b--c.

At each $\lambda$, observing that $\Sigma^{\lambda}$ must be a mixture of $\Sigma^{\text{TD}}$ and $\Sigma^{\text{BO}}$ with relative weights $1-\lambda$ and $\lambda$, we theorize
\begin{equation}\label{eq:qcvlambda}
\Sigma^{\lambda} = (1-\lambda)^{2}\Sigma^{\text{TD}}+\lambda^{2}\Sigma^{\text{BO}}+2\lambda(1-\lambda)\Sigma^{\text{TD,BO}},
\end{equation} 
with $\Sigma^{\text{TD,BO}}=\mathbb{E}\left[\left(\mathbf{u}^{\text{TD}}-\mathbb{E}\left[\mathbf{u}^{\text{TD}}\right]\right)\left(\mathbf{u}^{\text{BO}}-\mathbb{E}\left[\mathbf{u}^{\text{BO}}\right]\right)\right]$. The final ensemble sampling is done using $\Sigma^{\lambda}$ which leads to a much closer displacement profile to that of a $\lambda$-modulated AIMD ensemble. As shown in Fig. \ref{fig:displacements}d, $\Phi_{2}$-mixing generates unphysically large displacements, which $\Sigma$-mixing rolls back accurately toward the AIMD profile.


Once $\lambda$-dependent ensembles are obtained, their potential energies, $U_{\text{BO}}$ and $U_{\text{TD}}$, are calculated simply by Eq. \ref{eq:latticeenergy} using $\Phi^{\text{BO}}$ (all orders) and $\Phi^{\text{TD}}_{2}$ (2nd order only). In doing so, we replace DFT total energy calculations with a series of matrix operations. Note that any surrogate energy-calculating model could be used here if one wished. The anharmonic free energy, which is the difference between that of the TD-harmonic and BO potentials, is $F_{\text{anh}} = \int_{0}^{1} \langle U_{\text{BO}} - U_{\text{TD}} \rangle_{\lambda} d\lambda$ by Eq. \ref{eq:ti}. The total vibrational free energy is $F_{\text{vib}}=F_{\text{TD}}+F_{\text{anh}}$, where $F_{\text{TD}}$ is the harmonic free energy of the TD phonons, computed using Phonopy \cite{phonopy,phonopyreview}. The total Helmholtz free energy is $F=E_{\text{DFT}}+F_{\text{vib}}+F_{\text{e}}$, where $E_{\text{DFT}}$ is the DFT energy of the equilibrium structure and $F_{\text{e}}$ is the electronic free energy. $F_{\text{e}}$ is non-negligible for metals and thus included.


\begin{figure*}
\includegraphics[width=0.9\linewidth]{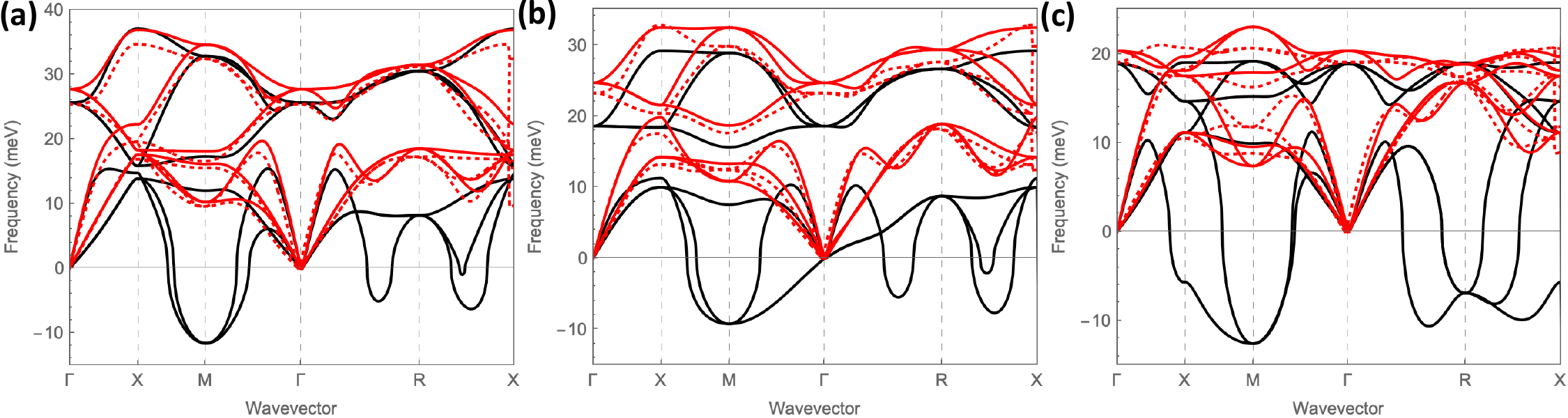}
\caption{\textbf{Temperature-dependent phonons of the B2-phases of the binary SMAs.} \textbf{a)} NiTi at 400 K, \textbf{b)} PdTi at 700 K, \textbf{c)} NiHf at 1400 K. Black solid dispersions are as obtained directly from original fitting to DFT. Red solid dispersions are computed using anharmonic renormalization. Red dotted dispersions are computed using TDEP.}
\label{fig:sma_phonon}
\end{figure*}

\begin{figure*}
\includegraphics[width=0.8\linewidth]{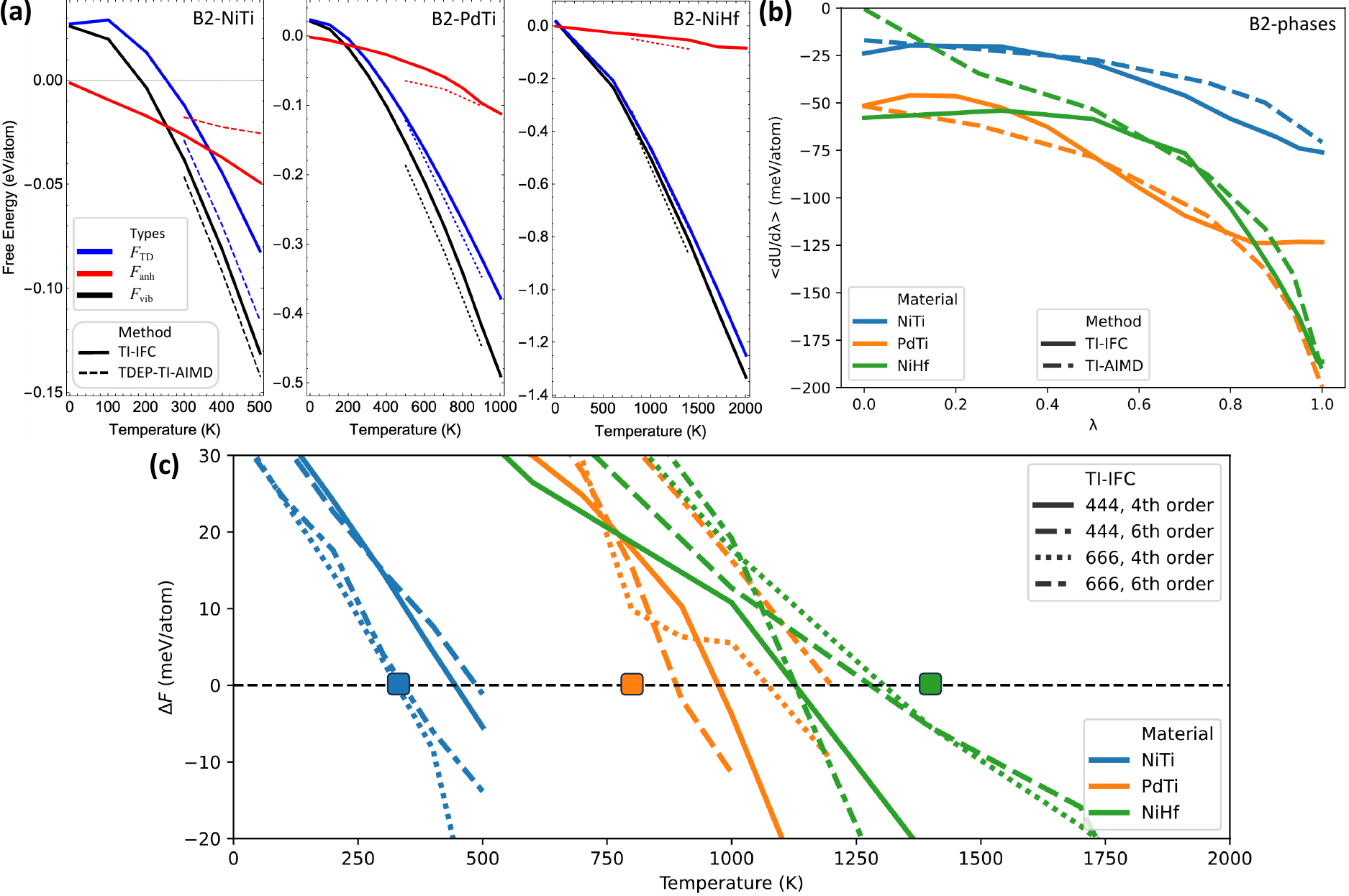}
\caption{\textbf{Free energy and phase transition of NiTi, PdTi, and NiHf.} The supercell size is $4\times4\times4$ for all three. \textbf{a)} Comparison of harmonic, anharmonic and total vibrational free energies of the B2-phases computed using TI-IFC and TDEP-TI-AIMD, and \textbf{b)} comparison of $\langle\frac{dU}{d\lambda}\rangle$ of the B2-phases at 400 K for NiTi (4th order), 800 K for PdTi (6th order), and 1400 K for NiHf (4th order), sampled and calculated with $\Sigma^{\lambda}$ and AIMD, using identical TD potential. \textbf{c)} $\Delta F$ between the B2-phases and the low-temperatures phases across levels of theory. The $\Delta F=0$ points indicate predicted $T_{c}$ values, and experimental $T_{c}$ values are marked with squares on the zero axis.}
\label{fig:sma_free_energy}
\end{figure*}


We begin by discussing in detail binary SMAs, NiTi, PdTi, and NiHf, and their martensitic phase transition. These make our primary test-beds due to their crucial aerospace applications (planetary rover wheels and aircraft actuators) and their involvement of very strong anharmonicity and complex potential energy surfaces. Phase transitions occur between the high-temperature austenite phases B2 and the various low-temperature martensitic phases: B19$^{\prime}$, B19, B33 for NiTi, PdTi, NiHf respectively. The B2-phases are computational challenging to describe due to the particularly strong anharmonicity. Their dynamical instabilities are shown in Fig. \ref{fig:sma_phonon} alongside the effective harmonic phonons obtained at high temperatures. Accurate vibrational free energies of these B2 phases are critical to describing the shape-memory transition, as has been demonstrated by the application of TI-AIMD \cite{tiaimdnitiHaskins,tiaimdnitiHaskins2,tiaimdbinarysmaHaskins,tiaimdnitiZW,tiaimdnitioffZW,tiaimdnitihfZW}. 

Fig. \ref{fig:sma_free_energy}a shows that the total $F_{\text{vib}}$ of the B2-phases computed using TI-IFC agree well with those computed using TDEP and TI-AIMD. Most strikingly in the case of NiTi, $F_{\text{TD}}$ notably differs depending on how TD phonons are obtained, but that difference is appropriately compensated by the $F_{\text{anh}}$ such that $F_{\text{vib}}$ is very close between the two methods. This is no coincidence but rather a direct, beneficial consequence of the theoretical foundation of TI, which describes the difference between the true potential energy and \textit{any} reference potential energy. This makes TI-IFC in principle agnostic to how TD phonons are generated and to some extent the details of TD phonons. TD phonons from one temperature could be used calculate $F_{\text{anh}}$ at another temperature.  TDEP \cite{tdep1,tdep2,tdep3}, SSCHA \cite{sscha1,sscha2,sschareview}, QSCAILD \cite{qscaild1,qscaild2} or any other method for generating TD phonnos can all be used. However, all of these methods invoke DFT calculations at every iteration of ensemble generation, by their standard protocol, so their use would likely negate the primary advantage of TI-IFC, speed. In any case, anharmonic IFC must be accurately calculated to ensure physical thermal ensembles and robustness of TI-IFC as a whole process. 

Fig. \ref{fig:sma_free_energy}b reveals more microscopic details of TI-IFC. When identical TD potential is used, TI-IFC generally tracks TI-AIMD closely in $\langle\frac{dU}{d\lambda}\rangle$ from $\lambda=0$ to $\lambda=1$ for all three B2-phases at the chosen temperatures. A 4th-order expansion is found to be sufficient for decent agreements with TI-AIMD in B2-NiTi and B2-NiHf. In contrast, B2-PdTi is considerably better described with a 6th-order expansion. This suggests that the anharmonic double well in B2-PdTi is more strongly modified by the 6th-order term than the other two materials. 

\begin{figure}
\includegraphics[width=1\linewidth]{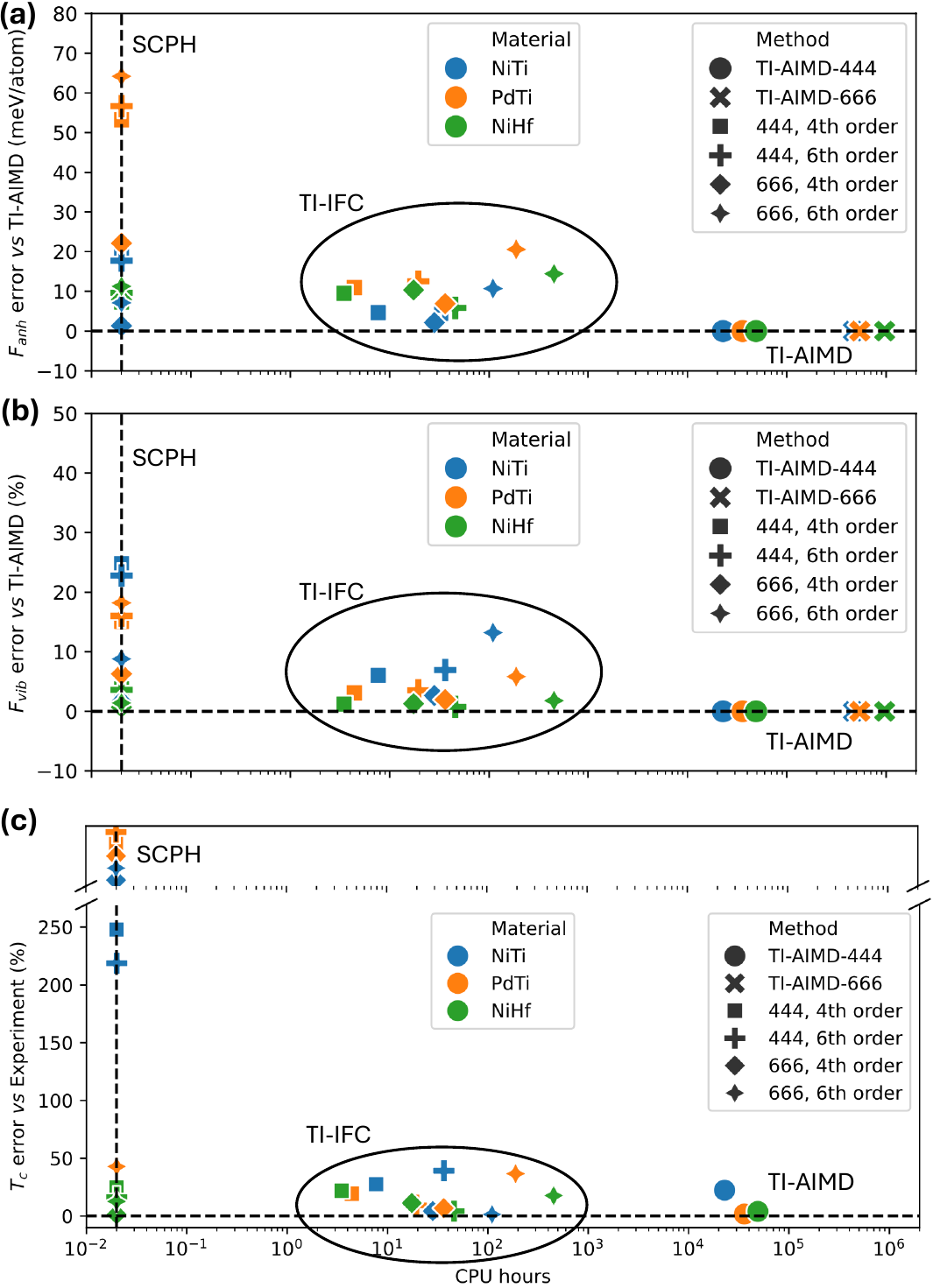}
\caption{\textbf{Computing hours vs accuracy of TI-IFC on binary SMAs.} CPU hours is defined for single-temperature calculation of the B2-phases (400 K for NiTi, 800 K for PdTi, 1400 K for NiHf). Results calculated using free energy corrections from perturbation theory are placed on the dashed vertical line representing the approximate CPU hours ($\sim10^{-2}$) for that calculation. The cases in which we did not find a phase transition within the temperature-range of study or extrapolation of results are plotted above the figure break. \textbf{a)} Hours vs error in anharmonic free energy from TI-AIMD in meV/atom. The same TD-harmonic potential is used for TI-IFC and TI-AIMD. \textbf{b)} Hours vs \%-error in the total vibrational free energy from TI-AIMD. \textbf{c)} Hours vs \%-error in $T_{\text{c}}$ from experimentally. TI-AIMD here is done with TDEP potentials as the reference.}
\label{fig:hours}
\end{figure}

The free energy difference ($\Delta F$) between the competing phases and the predicted $T_{\text{c}}$ by TI-IFC are plotted in Fig. \ref{fig:sma_free_energy}c. The predicted $T_{\text{c}}$ using the minimal setting ($4\times4\times4$ supercells, 4th order) are about 440 K for NiTi, 960 K for PdTi, and 1130 K for NiHf, which are within 25\% error from experimental values, 345 K, 810 K, 1420 K, respectively. Accuracy improves to within 10 \% error if larger $6\times6\times6$ supercells are used, where the predicted $T_{\text{c}}$ are 330 K for NiTi, 860 K for PdTi, and 1360 K for NiHf. The same is not strictly true of more complex lattice models involving higher orders, as exemplified by some cases of 6th-order expansion shown in Fig. \ref{fig:hours}. This reflects that, for given training set, prescription of a supposedly more physical model with more IFC parameters to be solved for could degrade the numerical solution and the model quality. Quantitative differences put aside, phase transition is predicted under all settings for all materials. 

The speed-up of TI-IFC over TI-AIMD is about 4 orders of magnitude depending on the model complexity. Note that, once compounded over multiple temperatures and phases, the absolute speed-up would be far greater than shown on scale. The speed-up comes at the expense of generally within 10 meV/atom error in $F_{\text{anh}}$, which translates to within 5\% error in the accuracy of $F_{\text{vib}}$, in most cases. Error in predicted $T_{c}$ is generally within 20\% of experimental values. While essentially cost-free, free energy correction from perturbation theory is not sufficiently accurate for the SMAs. For NiTi and PdTi, it fails to predict phase transitions at all in most cases. Only in NiHf where $F_{\text{TD}}$ dominates does the perturbative correction predict phase transition.

\begin{figure*}
\includegraphics[width=0.65\linewidth]{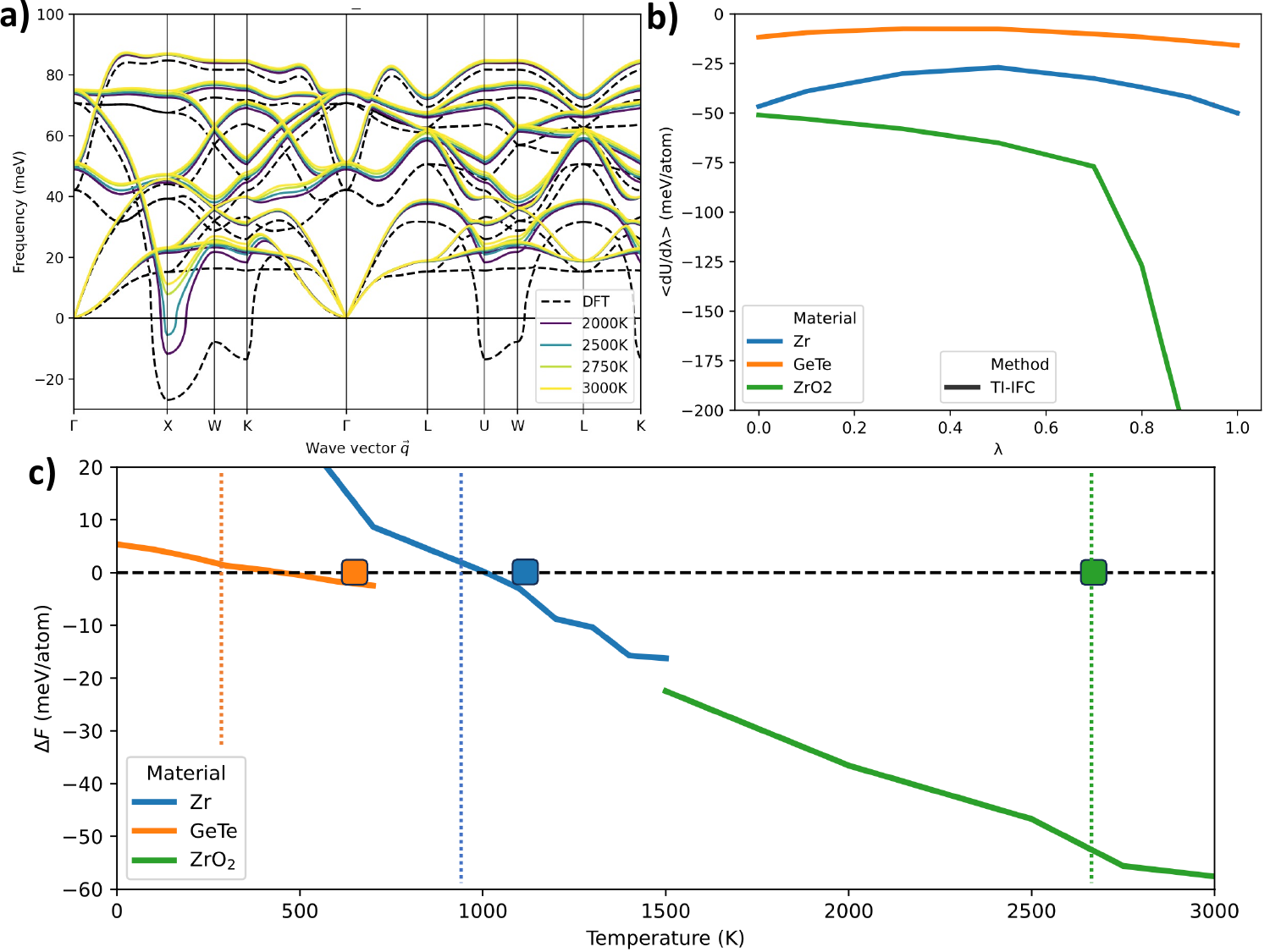}
\caption{\textbf{TI-IFC results on other systems, Zr, GeTe, and ZrO$_{2}$.} \textbf{a)} The temperature-dependent phonons of cubic ZrO$_{2}$ ($5\times 5\times 5$ supercell). \textbf{b)} The $\langle\frac{dU}{d\lambda}\rangle$ trace for at 1100 K, 600 K, and 2750K respectively for BCC-Zr, cubic GeTe and cubic ZrO$_{2}$ (high-temperature phases) \textbf{c)} $\Delta F$ between the high-temperature phases and low-temperatures phases HCP-Zr, rhombohedral GeTe, tetragonal ZrO$_{2}$. Temperatures where dynamical stability is achieved is marked with dotted vertical lines. Experimental $T_{c}$ values are marked with squares on the zero axis. For GeTe and Zr, the predicted $T_{\text{c}}$ is where $\Delta F=0$. For ZrO$_{2}$, the predicted $T_{\text{c}}$ is where the phonons become stable ($\approx 2600$ K) since $\Delta F<0$ immediately.}
\label{fig:others}
\end{figure*}

We apply TI-IFC to other dynamically unstable systems, namely ZrO$_{2}$, GeTe, and Zr. For these materials, Eq. \ref{eq:latticeenergy} for fitting is expanded up to 4th order for all phases involved, the minimally required setting. The phonon dispersions and $\langle\frac{dU}{d\lambda}\rangle$ values of the high-temperature phases are shown in Fig. \ref{fig:others} as well as $\Delta F$.

The tetragonal-to-cubic transition in ZrO$_{2}$ is a challenging phenomenon to study due to not only the highly anharmonic potential in the cubic phase but also the extremely high $T_{c} \approx 2600$ K, only marginally below the melting point ($\approx3000$ K) \cite{zro2structure}. The phonon dispersion of the cubic phase, shown in Fig. \ref{fig:others}a, has imaginary modes that become stable between 2500K and 2750K, or at about 2600 K if interpolated. Observation of structural fluctuations during MD using MLFF has led to $T_{c}$ of 2450 K \cite{mdmlffzro21} and 2750 K \cite{mdmlffzro22} depending on the method of force-field training. In another recent study, free energies by perturbative corrections were unsuccessful in predicting a phase transition as the tetragonal phase had lower free energy than the cubic phase up to 3000 K, suggesting perhaps that anharmonicity up to 4th order is not sufficient for describing this phase transition \cite{alamodezro2}. Our TI-IFC result predicts that the cubic phase theoretically has the lower free energy immediately as it becomes dynamically stable around 2600 K. The conclusion is then $T_{c}\approx2600$ K, squarely in the ballpark of the values determined by both experiments and MD-MLFF computations. This suggests that the 4th-order anharmonicity does provide a sufficient explanation for this phase transition. Despite the apparent success, we note the likelihood of higher quantitative uncertainty with ZrO$_{2}$ calculations owing to the extremely high temperature at play. Aside from the inherent approximations of our method, we do not consider thermal expansion, which can be of substantial effect at such high temperatures.

GeTe is a ferroelectric \cite{getespintexture,geteferroelectric,getebite3ferroelectric} as well as a good thermoelectric \cite{getejoule,getepnas,gesbinte} exhibiting a rhombohedral-to-cubic phase transition, whose experimentally reported critical temperature is rather wide-ranged at $T_{c}\approx650\pm100$ K \cite{getephasetransition1,getephasetransition2,getephasetransition3}. The high-temperature cubic phase has unstable optical modes at and near the $\Gamma$-point as seen in SI Fig. S2, which attain dynamical stability by 300 K. GeTe too has been studied via MD-MLFF which predicted a $T_{c}=630$ K \cite{mdmlffgete}. TI-IFC free energy calculation underpredicts at $T_{c}\approx450$ K. As seen in Fig. \ref{fig:others}e, the scale of $\Delta F$ between the two GeTe phases is tiny, varying only by 8 meV/atom over an 800 K temperature range. This is in the vicinity of inherent numerical uncertainties of DFT. An error of 5 meV/atom in would, for instance, sway $T_{c}$ by nearly 500 K. $T_{c}$ here is thus particularly sensitive to even slight changes in not only $F_{\text{vib}}$ but also $E_{\text{DFT}}$. If PBE and not PBEsol is used to calculate $\Delta E_{\text{DFT}}$, the predicted $T_{\text{c}}$ of GeTe would be 700 K. In hindsight, the wide range of experimental $T_{c}$ is perhaps indicative of small $\Delta F$.

Zr is a classic elemental metal \cite{latticeinstabilitiesinmetallicelements} afflicted by dynamical instability. It is known to undergo an HCP-to-BCC transition around 1134 K \cite{latticedynamicszr,latticedynamicsbcczr}. The high-temperature BCC-phase has a pronounced imaginary modes at the $N$-point as seen in SI Fig. S2, which attains stability at around 700 K. TI-IFC predicts $T_{c}\approx1030$ K, close to the experimental $T_{c}$. This phase transition has also been subject to a MD-MLFF study \cite{mdmlffzr}, which determined a $T_{\text{c}}\approx 1000$ K from observing structure fluctuations.




We demonstrate a new approach to thermodynamic integration using only interatomic force constants capable of offering anharmonic free energy corrections at orders-of-magnitude higher computational throughput capacity than the traditional route via AIMD. The flowchart for this method is provided in Fig. \ref{fig:flowchart} The validity of our approach at the theoretical level rests on the assumption that force constants up to an arbitrary order constitute a good description of the true lattice potential at temperatures relevant to phase transition of interest. This is necessary to ensure that 1) the configurational ensemble is a reasonable approximation of the true thermal ensemble explored by AIMD; 2) Eq. \ref{eq:latticeenergy} is a reasonable approximation of the true potential energy. In the systems hereby tested, the said assumption appears to hold to a large extent, and the approach as a whole describes their phase transition behaviors well. 

We stress that TI-IFC is a method meant primarily to be efficient whose accuracy is systematically improvable and ultimately reasonable. For instance, while 4th-order expansion of IFC is expected to provide reasonable accuracy in most use cases, there is no general rule dictating what order of expansion or supercell size, cutoff radius, training set size and ensemble size, constitute ``sufficient accuracy" for one's purpose and target of study. Multiple settings maybe tested at one's own discretion to the point of building confidence in the accuracy needed for the intended purpose, whether it is large-scale screening or deep-dive into a specific system. Even when emphasis is placed on accuracy over speed, this method should be a couple of orders of magnitude more expedient than TI-AIMD at least for relatively simple solids, and further speed-up can come via more efficient programming. The combination of general applicability, accuracy and efficiency of this method is especially well-suited for high-throughput implementations for materials discovery and phase diagram construction across a large chemical space.

\acknowledgements
This work was funded by the Transformational Tools and Technologies (TTT) project of Aeronautics Research Mission Directorate (ARMD) at National Aeronautics and Space Administration (NASA).




\bibliography{references}

\end{document}


\title{Supplementary Information}
	\author{Junsoo Park}
	\email{junsoo.park@nasa.gov}
	\affiliation{\kbrnasaames}
    \author{Zhigang Wu}
    \affiliation{\nasaames}
    \author{John W. Lawson}
	\affiliation{\nasaames}
	\date{\today} 
	\maketitle

\section{DFT Calculations and IFC Fitting}

We generate the training structures for IFC fitting by performing by randomly displacing the atoms in the relaxed supercells. For the high-temperature phases, we perturb the supercell by $0.01\sim0.5$ {\r A}. Ten total training structures are used for binary shape-memory alloys. A plane-wave cutoff of 600 eV is used for all DFT calculations. Cutoff distance in Angstroms from 2nd to 6th order clusters are [6.0, 5.5, 5.0, 4.0, 4.0] for $4\times4\times4$ B2-NiTi, [6.5, 5.5, 5.3, 4.5, 4.5] for B2-PdTi, and [6.5,6.3,5.4,4.5,4.5] for $4\times4\times4$ B2-NiHf. Recursive feature elimination is used for IFC fitting. 

One note - for every material and supercell size, we use the same training set for 4th-order and 6th-order fittings. This means that, because a 6th-order expansion has many more IFC parameters than a 4th-order expansion, it has relative shortage of training data compared to the latter. While a higher-order expansion captures more physics and thus is more accurate in theory, due to the smaller amount of training data per parameter, it does not guarantee higher accuracy. The only certain way to systematically improve accuracy is to expand to higher orders while enlarging the training data commensurately. We do not do this for this study.

\section{Temperature-dependent Phonons}

\begin{figure*}
\includegraphics[width=1\linewidth]{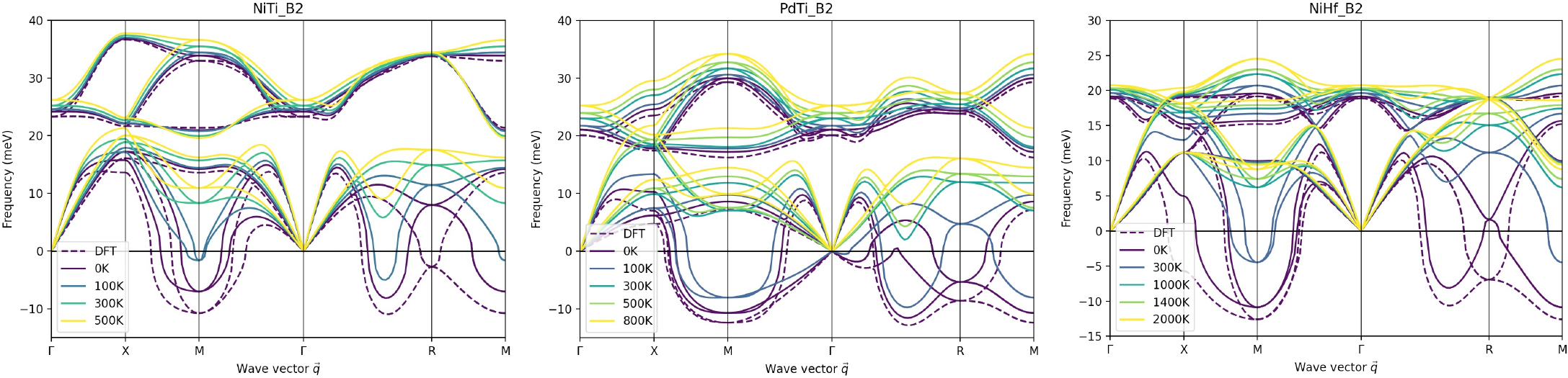}
\caption{\textbf{Phonon dispersions of the low-temperature phases of the binary SMAs.} In order, B19$^{\prime}$-NiTi, B19-PdTi, B33-NiTi. IFC has been fit up to 4th order.}
\label{fig:sma_phonon_supp}
\end{figure*}

\begin{figure*}
\includegraphics[width=0.7\linewidth]{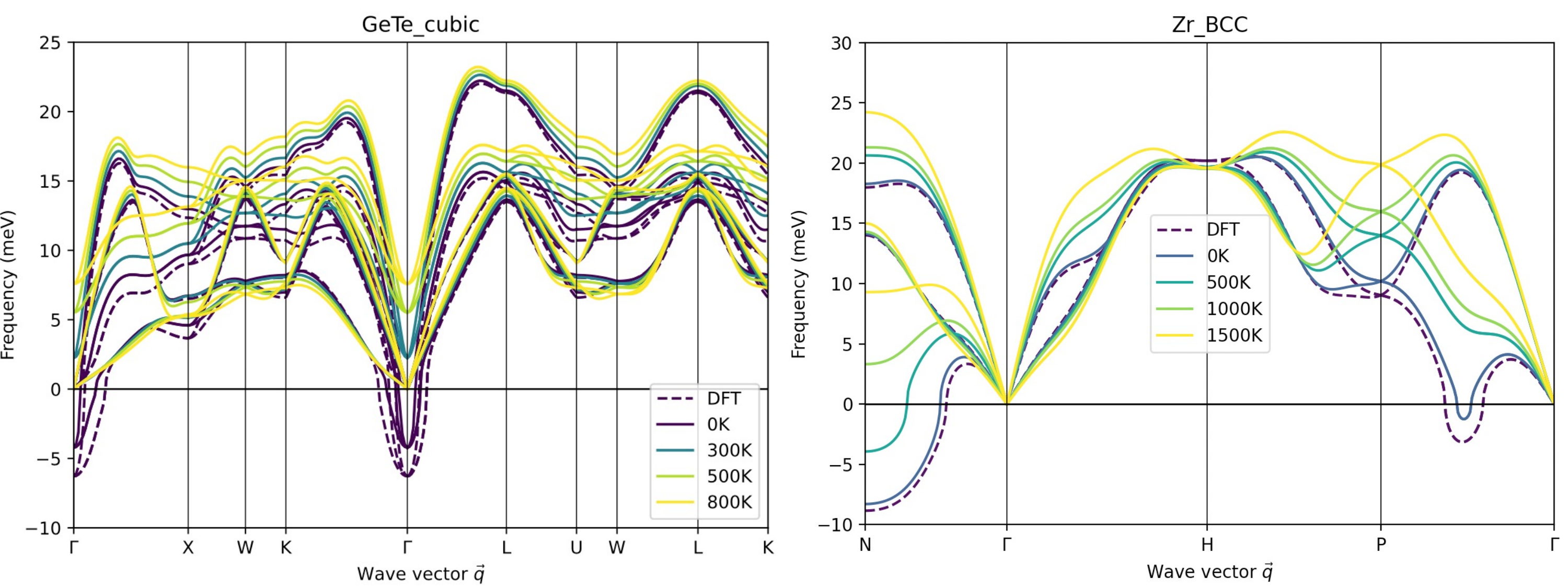}
\caption{\textbf{Phonon dispersions of BCC-Zr and cubic GeTe.} IFCs have been fit up to 4th order. $5\times5\times5$ supercells are used. }
\label{fig:getezr_phonon_supp}
\end{figure*}

Detailed temperature-dependent phonon dispersion of the B2-phases of NiTi, PdTi and NiHf is provided in Fig. \ref{fig:sma_phonon_supp}. Temperature-dependent phonon dispersion of the BCC-Zr and cubic GeTe are shown in Fig. \ref{fig:getezr_phonon_supp}. These high-temperature phases all exhibit imaginary modes whose free energy requires anharmonic free energy calculation for accurate treatment. These are the primary targets of applying TI-IFC.

\begin{figure*}
\includegraphics[width=1\linewidth]{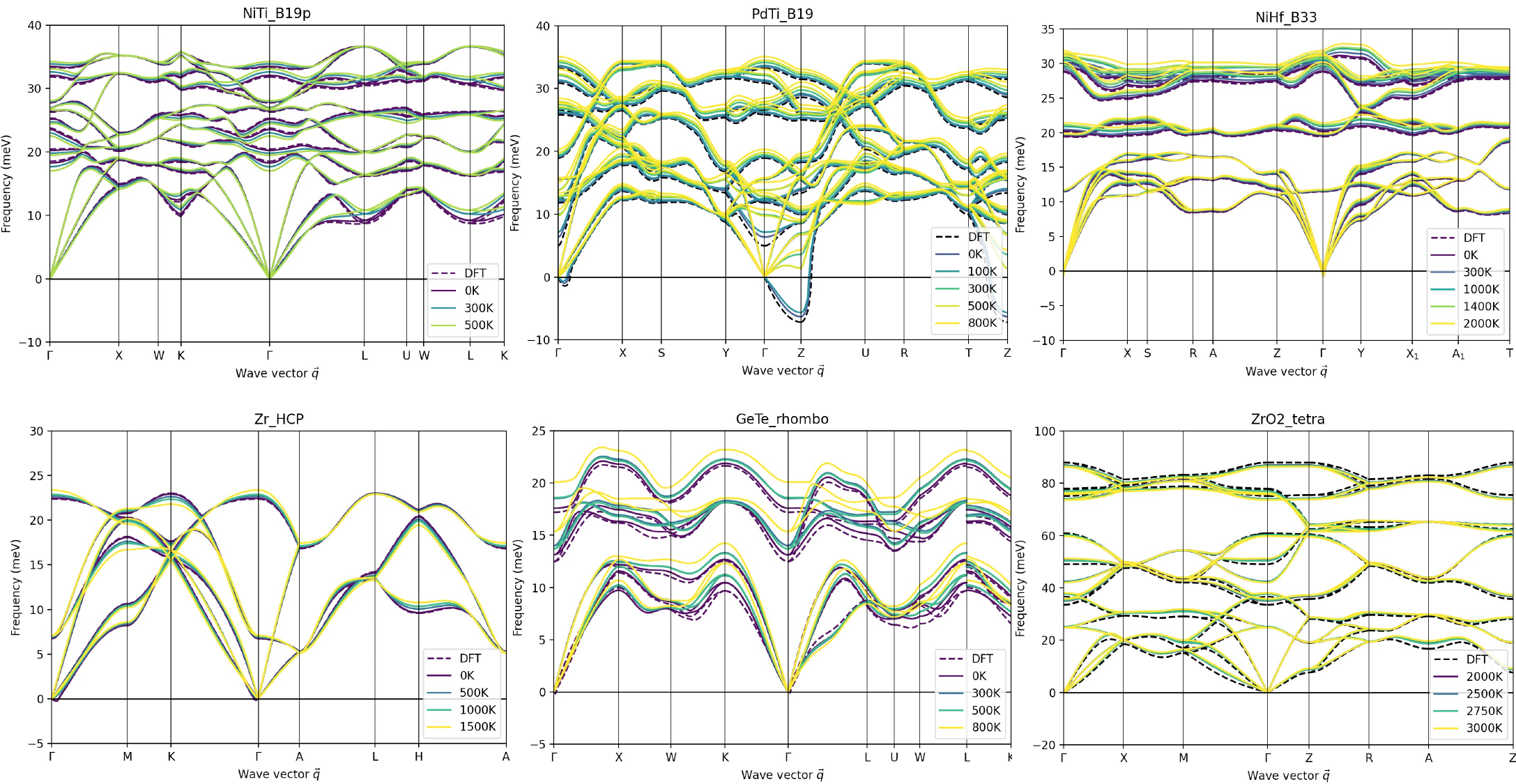}
\caption{\textbf{Phonon dispersions of HCP-Zr, rhombohedral GeTe, and tetragonal ZrO$_{2}$.} IFC has been fit up to 4th order.}
\label{fig:others_phonon_supp}
\end{figure*}

Anharmonic phonon renormalization and TI-IFC is performed on the low-temperature Martensitic phases of the SMAs and the low-temperature phases of Zr, GeTe, and ZrO$_{2}$ as well. The phonno dispersions are shown in Fig. \ref{fig:others_phonon_supp}. Cutoff distances up to 4th order for these phases are  [5.6,5.0,4.5], [6.2,5.0,4.5], and [6.2,5.0,4.5], respectively for B19$^{\prime}$-NiTi, B19-PdTi, and B33-NiHf. Other than in B19-PdTi, we do not find any imaginary modes among these. For the cases other than B19-PdTi, neither temperature-dependent phonons nor anharmonic free energy would typically even be calculated because anharmonic effects are small, and quasiharmonic free energy is sufficient. However, for consistency's sake, we do TD-phonon and TI-IFC calculations for all of the low-temperature phases.

\section{Thermal Ensembles}

The number of displaced configurations sampled for thermal ensembles scales with temperature and $\lambda$ as $N=N_{0}(1+5\lambda)(1+\frac{T}{100})$. This scaling accounts for the larger vibrational degrees of freedom at higher temperatures (square-displacement tends to increase linearly with temperature) and under more anharmonic potentials (larger $\lambda$). We set $N_{0}=3$ for the B2-phases in this study but it is an arbitrary user input.

\begin{figure*}
\includegraphics[width=0.85\linewidth]{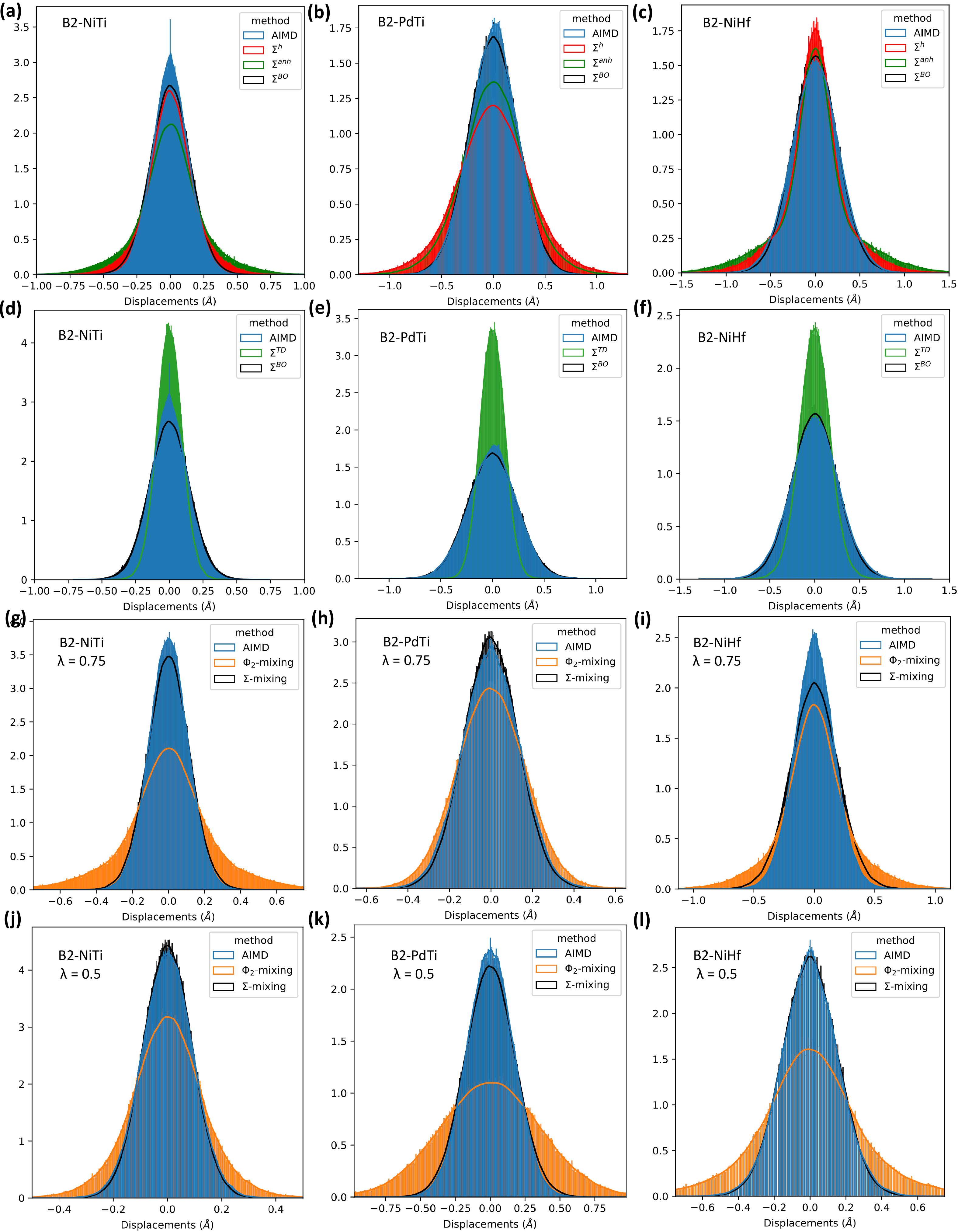}
\caption{\textbf{Detailed comparison of B2-phase displacement profiles of ensembles generated by various methods.} The $\Sigma^{\text{BO}}$-ensemble naturally corresponds to $\lambda=1$. The $\Sigma^{\text{TD}}$-ensemble corresponds to $\lambda=0$ in the TI context, but is generally used in the literature to describe the effective dynamics at finite temperatures. The $\Sigma^{\text{h}}$-ensemble encodes only the bare harmonic IFC fitted to DFT. The $\Sigma^{\text{anh}}$-ensemble is related to $\Phi_{2}^{\text{anh}}$. ``$\Phi$-mixing" indicates $\lambda$-dependent ensembles constructed from $\Phi^{\lambda}_{2}$, which fails when $\Phi^{\lambda}_{2}$ exhibits imaginary modes. Finally, ``$\Sigma$-mixing" indicates $\lambda$-dependent ensembles constructed from $\Sigma^{\lambda}$ of Eq. 7 in the main text. AIMD is run for each $\lambda$ case and serves as the benchmark.
By column,
\textbf{a,d,g,i)} B2-NiTi at 400 K using $4\times4\times4$ a supercell and 4th order expansion. \textbf{b,e,h,k)} B2-PdTi at 800 K using $4\times4\times4$ a supercell and 6th order expansion. \textbf{c,f,i,l)} B2-NiHf at 1400 K using $4\times4\times4$ a supercell and 4th order expansion.
By row,
\textbf{a-c)} Ensembles sampled from $\Sigma^{\text{BO}}$ reflecting the Born-Oppenheimer potential reconstructed from harmonic and anharmonic potentials, which individually are partial descriptions, successfully approximates AIMD displacement profiles. These correspond to $\lambda=1$ cases, by default. \textbf{d-f)} Ensemble sampling from $\Sigma^{\text{BO}}$ ($\lambda=1$) better approximates AIMD displacement profiles than $\Sigma^{\text{TD}}$ ($\lambda=0$). \textbf{g-i)} Comparison of displacement profiles of $\Sigma^{\lambda}$ and AIMD ensembles at $\lambda=0.75$. \textbf{j-l)} Comparison of displacement profiles of $\Sigma^{\lambda}$ and AIMD ensembles at $\lambda=0.5$.
}
\label{fig:displacements_supp}
\end{figure*}

Detailed comparison of displacement profiles of ensembles generated via the methods of this work and AIMD are provided in Fig. \ref{fig:displacements_supp} for all three binary SMAs. In all cases, $\Sigma^{\text{BO}}$ and $\Sigma^{\lambda}$ approximate the AIMD ensemble very closely, and better than any other sampling method. Fig. 1a in the main text suggest that the bare harmonic potential would generate a more dispersed displacement profile than the anharmonic potential. This however is only true of in the directions of instability. In all other directions where the harmonic well is upright, it may generate tighter displacement profile. Therefore, the overall displacement profile may be more spread under harmonic or anharmonic potentials.

The cross-covariance terms that appear in Eqs. 6 and 7 in the main text are in effect 0, by both theory and numerical confirmation. $\Sigma^{\text{h,anh}}=0$ due to the independence of harmonic and anharmonic potentials. $\Sigma^{\text{TD,BO}}=0$ is perhaps less obvious since $\Sigma^{\text{TD}}$ and $\Sigma^{\text{BO}}$ are constructed from the same ingredients, but the underlying distributions are ultimately independent of one another.

\begin{figure*}
\includegraphics[width=1\linewidth]{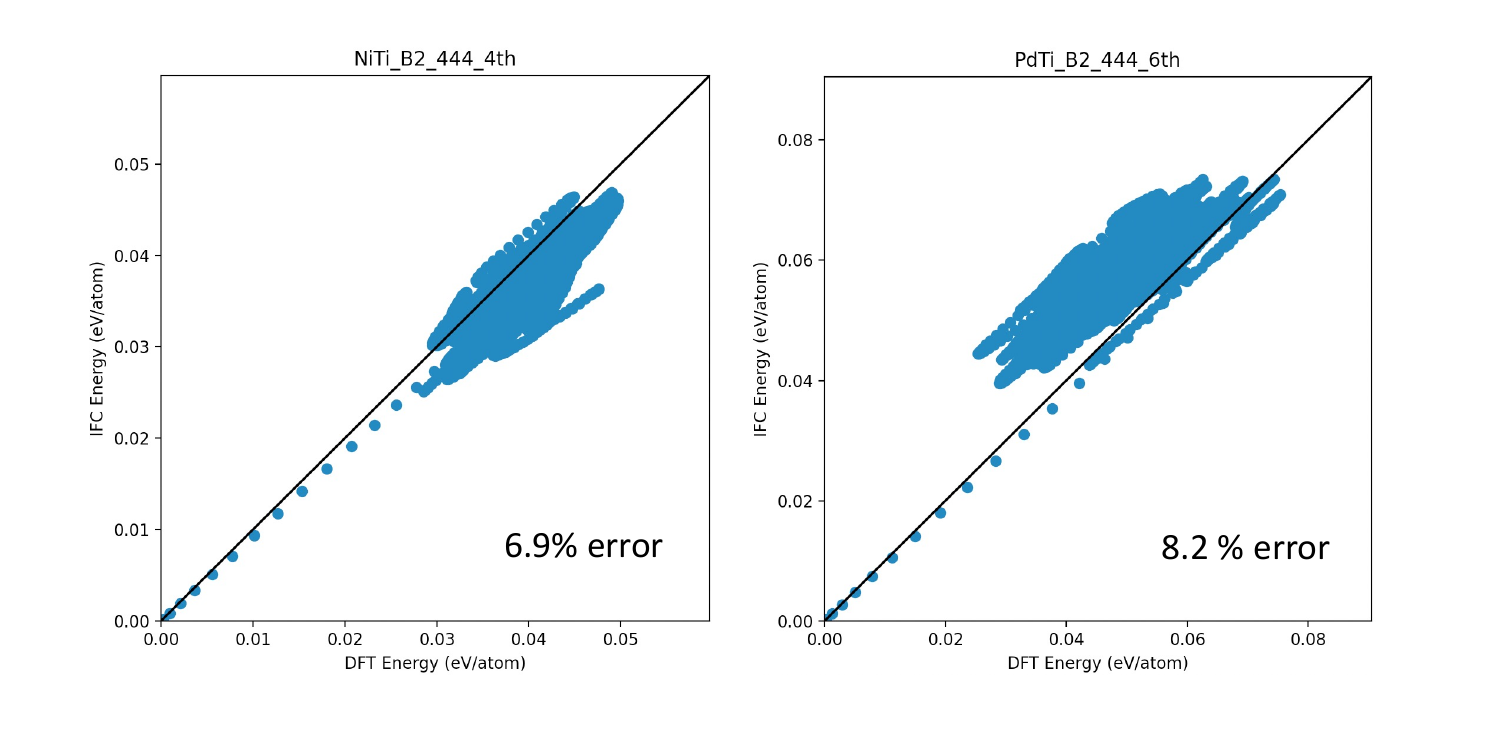}
\caption{\textbf{Energies of AIMD ensemble configurations calculated using IFC vs DFT.} The values are referenced to the baseline DFT energy of the equilibrium structure. Supercells used are $4\times4\times4$ for both. 4th-order expansion is used for NiTi and 6th-order expansion is used fro PdTi.}
\label{fig:energy_comparison}
\end{figure*}

IFCs are used not only to generate QCVs and sample ensembles, but also to predict ensemble potential energy through Eq. 1 of the main text. To assess their fidelity as energy predictor, we calculate and compare energies of AIMD ensemble using both DFT and IFC and compare them in Fig. \ref{fig:energy_comparison}. Relative error of IFC energy to DFT energy is $6\sim9\%$ for B2 NiTi (4th-order expansion) and PdTi (6th-order expansion).

\section{Thermodynamic Integration}

The default $\lambda$-grid used for all materials at each temperature is [0,0.1,0.3,0.5,0.7,0.8,0.9,0.95,1.0]. This is reflective of the general observation that $\frac{dU}{d\lambda}$ changes slowly at low $\lambda$ and more rapidly at high $\lambda$.

Due to the statistical nature of the sampling strategy, it is not impossible that a structure of extremely large displacements and high energy is sampled out of proportion. Albeit rarely occurring, such an outliar could potentially throw off the ensemble average at a $\lambda$-point and ultimately the final free energy to unphysical extents. To suppress this effect from corrupting $\lambda$-integration, we apply a safety filter that excludes a $\lambda$-point whose $\langle\frac{dU}{d\lambda}\rangle$ value exhibits excessive deviation. Specifically, we exclude any $\langle\frac{dU}{d\lambda}\rangle$ that deviate from neighboring values within $\lambda\pm0.1$ by more than 100 meV/atom.

\section{TI-AIMD}

For benchmarking purpose, we run TI-AIMD calculations using VASP with two types of temperature-dependent harmonic IFCs: ones generated by TDEP and ones used for TI-IFC generated by anharmonic phonon renormalization such that the TD harmonic free energy is kept identical. The chosen temperatures are 400 K for NiTi, 1400 K for NiHf, and 800 K for PdTi, which are near their respective critical temperatures. 

We adopt the same approaches and settings as detailed in Ref. \onlinecite{tiaimdnitiZW}. In summary, the thermodynamic average for a fixed $\lambda$ value is obtained from the canonical $NVT$ AIMD simulations with the mixed potential $U_{\lambda}$ for 4 to 20 ps, to ensure convergence. $\lambda = 0.0, 0.25, 0.50, 0.75, 1.0$, $\lambda = 0.0, 0.25, 0.50, 0.75, 0.875, 1.0$, and $\lambda = 0.0, 0.25, 0.50, 0.75, 0.875, 0.9375, 1.0$ are used for NiTi, PdTi and NiHf, respectively; the time step is 2.0 fs, and the friction factor for the Langevin thermostat is 100 fs. The PBE exchange-correlation functional and the projector augmented wave (PAW) psuedopotentials are used in DFT calculations. When computing free energy from TDEP-computed phonons, the plain PBE functional is used for the entire process. When computing free energy from phonon renormalization results for a direct comparison with TI-IFC, PBEsol is used to maintain consistency. The valence of Ni, Ti, Pd and Hf atoms include 3d$^{9}$4s$^{1}$, 3d$^{2}$4s$^{2}$, 4d$^{10}$ and 5d$^2$6s$^2$ electrons, respectively. 


\section{Perturbative Correction to Free Energy}

For comparison, we also calculate anharmonic free energy correction using expressions obtained from the self-consistent phonon theory \cite{cowleylatticedynamics,alamodesrtio3}. In that approach, anharmonic free energy correction can be calculated for 4th-order anharmonicity (considered first-order correction) and-3rd-order anharmonicity (considered second-order correction). The correction due to the 4th-order anharmonicity for TD phonons, the so-called ``loop" correction, is 
\begin{equation}\label{eq:scph}
F_{\text{anh}} = -\frac{1}{4}\sum_{\lambda\mathbf{q}} \frac{(\omega_{\lambda\mathbf{q}}^{2}(T)-\mathbb{C}^{\dagger}\omega_{\lambda\mathbf{q},\text{h}}^{2}\mathbb{C})}{\omega_{\lambda\mathbf{q}}(T)}\left(n_{\lambda\mathbf{q}}(T)+\frac{1}{2}\right),
\end{equation}
where $\mathbb{C}$ is the unitary transformation matrix between the TD and bare harmonic eigenvectors, and subscript ``h" stands for bare harmonic originally obtained through DFT. This is the correction term we compute and compare in Fig. 5 of the main text. The correction term due to the 3rd-order anharmonicity (``bubble correction") is more complex to calculate \cite{improvedsc1968,alamodescf3}, and we omit it as such. The benefit of Eq. \ref{eq:scph} is that they take essentially no time to compute once TD phonons are obtained. However, they are insuficient accounts of anharmonic free energy, as evidenced in Fig. 5, when the underlying potential is strongly anharmonic.

\section{Electronic Free Energy}

Electronic free energy is accounted for the binary SMAs which are metals. $F_{\text{e}}$ is computed from the Fermi-Dirac distribution $f(E)$ and the electronic density of states $D(E)$ calculated using the tetrahedron method \cite{tetrahedron}: 
\begin{equation}\label{eq:electronic}
F_{\text{e}} = -k_{\text{B}}T\int D\left[f\text{log}(f) + (1-f)\text{log}(1-f)\right] dE
\end{equation}

\bibliography{references}